\documentclass[pre,superscriptaddress,showpacs,preprint]{revtex4}
\usepackage{amsmath,graphicx,amssymb,amsfonts}

\begin{document}

\title{The Physics of Granular Mechanics}
\author{Yimin Jiang}
\affiliation{Central South University, Changsha, China 410083}
\author{Mario Liu}
\affiliation{Theoretische Physik, Universit\"{a}t T\"{u}bingen, 72076 T\"{u}bingen,
Germany}
\date{\today }

\begin{abstract}{The {\em hydrodynamic} approach to a
continuum mechanical description of granular behavior
is reviewed and elucidated. By considering energy and
momentum conservation simultaneously, the general
formalism of {\em hydrodynamics} provides a systematic
method to {derive} the structure of constitutive
relations, including all gradient terms needed for
nonuniform systems. An important input to arrive at
different relations (say, for Newtonian fluid, solid
and granular medium) is the energy, especially the
number and types of its variables.
\\
Starting from a careful examination of the physics
underlying granular behavior, we identify the
independent variables and suggest a simple and
qualitatively appropriate expression for the granular
energy. The resultant hydrodynamic theory, especially
the constitutive relation, is presented and given
preliminary validation.}
\end{abstract}\maketitle

\section{Introduction}
When unperturbed, sand piles persist forever,
demonstrating in plain sight granular media's ability
to sustain shear stresses -- an ability that is
frequently considered the defining property of solids.
On the other hand, when tapped, the same pile quickly
degrades, to form a layer (possibly a monolayer) of
grains minimizing the gravitational energy. This is
typical of liquids. The microscopic reason for this
dichotomy is clear: The grains are individually (and
ever so slightly) deformed if buried in a pile, which
is what sustains the shear stress. When tapped, the
grains jiggle and shake, and briefly loose contact
with one another. This is why they get rid of some of
their deformation -- which shows up, macroscopically,
as a gradual lost of the static shear stress and a
continual flattening of the pile.~\footnote{In the
usual picture, force chains consisting of infinitely
rigid grains is what sustains shear stresses. This
does not contradict the above scenario, it is just a
different description of the same circumstance: With
contacts that are Hertzian (or Hertz-like), grains are
infinitely soft at first contact, irrespective of how
stiff the bulk material is, since very little material
is then being deformed. There is therefore,
realistically speaking, always some granular
deformation and elastic energy present in any force
chain. Now, because grains get rapidly stiffer when
being compressed further, the displacement is small,
and infinite rigidity is frequently a good
approximation. Yet it is the loss of this tiny
deformation by tapping that is the cause for the
flattening of the pile.}

When sand is being sheared at a constant rate, both solid
and fluid behavior are operative. First, the grains are
being deformed, increasing the shear stress as any solid
would. Second, the same shear rate also provokes some
jiggling, just as if the grains were lightly
tapped.~\footnote{``Jiggling" is used throughout, for
random motion of the grains, large or small, that occurs
when they are rearranging. Sometimes, words such as
wiggle, creep, or crawl may be more appropriate.} This
leads to a fluid-like relaxation of the shear stress --
the larger the shear rate, the stronger the jiggling, and
the quicker the relaxation. Note the reason why loading
and unloading give different responses (called {\em
incremental nonlinearity}~\cite{Kolym-1,Kolym-2}): When
being loaded, the solid part of granular behavior
increases the stress, while the fluid part decreases it.
During unloading, both work in the same direction to
reduce the stress.

This entangled behavior, we suspect, lies at the heart of
the difficulty modeling sand macroscopically. In
addition, there is a ``history-dependence" of granular
behavior that, being experimentally obvious but
conceptually confused and ill-defined, further perplexes
the modeler. Obviously, if sand can be characterized, as
do other systems, by a {\em complete set of state
variables}, any history-dependence only indicates that
the experiments were run at different values of these
variables. This is what we believe happens.

The  {\em hydrodynamic theory} is a powerful approach
to continuum-mechanical description (or macroscopic
field theory), pioneered by Landau~\cite{LL6} and
Khalatnikov~\cite{Khal} in the context of superfluid
helium. Bei considering energy and momentum
conservation simultaneously, and combining both with
thermodynamic considerations, this approach is capable
of cogently deducing, among others, the proper
constitutive relation. Hydrodynamics~\cite{hydro-1}
has since been successfully employed to account for
many condensed systems, including liquid crystals
~\cite{liqCryst-1,liqCryst-5}, superfluid
$^3$He~\cite{he3-1,he3-3,he3-6},
superconductors~\cite{SC-1,SC-2,SC-3}, macroscopic
electro-magnetism~\cite{hymax-1,hymax-2,hymax-4} and
ferrofluids~\cite{FF-2,FF-3,FF-8,FF-9}. Transiently
elastic media such as polymers are under active
consideration at
present~\cite{polymer-1,polymer-3,polymer-4}.

Two steps are involved in deriving the theory
hydrodynamically, the first specifies the theory's
{\em structure}: Being a function of the state
variables, the energy itself is not independent.
Nevertheless, the form of the energy density
$w(s,\rho)$ is left unspecified in this first step,
and the differential equations are given in terms of
the energy density $w$, its variables and conjugate
variables. [Conjugate variables are the derivatives of
the energy with respect to the variables, say
temperature $T(s,\rho)\equiv\partial w/\partial s$ and
chemical potential $\mu(s,\rho)\equiv\partial
w/\partial\rho$ for $s,\rho$, the entropy and mass
density]. In a continuum theory, a number of transport
coefficients [such as the viscosity $\eta(s,\rho)$ or
the heat diffusion coefficient $\kappa(s,\rho)$] are
needed to parameterize dissipation and entropy
production. Neither is their functional dependence
specified.

A theory is unique and useful, of course, only when its
energy and transport coefficients are made specific, in a
second step. This division is sensible, because the first
step is systematic, the second is not. The first starts
with clearly spelt-out assumptions based on the basic
physics of the system at hand, which is followed by a
derivation that is algebraic in nature, and hence rather
cogent. The second step is a fitting process -- one looks
for appropriate expressions, by trial and error, for a
few scalar functions that, when embedded into the
structure of the theory, will yield satisfactory
agreement with the many experimental data.

Starting from the physics of granular deformation and its
depletion by jiggling, we have identified the variables
and derived the structure of the equations governing
their temporal evolution~\cite{JL6}, calling it {\sc
gsh}, for granular solid hydrodynamics. But our second
step is not yet complete, and some proposed functional
dependencies are still tentative. The expression for the
energy appears quite satisfactory, but our notion of the
transport coefficients is still vague. Our final goal is
a transparent theory with a healthy mathematical
structure that is capable of modeling sand in its full
width of behavior, from static stress distribution, via
elastoplastic deformation~\cite{elaPla-1,elaPla-2}, to
granular flow property at higher
velocities~\cite{Haff-1,Haff-2,chute-2,chute-3}.

\section{Granular State Variables}
In this section, we determine the complete set of
granular variables starting from the elementary
physics of granular deformation and its depletion by
jiggling.

\subsection{The Elastic Strain}

If a granular medium is sheared, the grains jiggle,
roll and slide, in addition to being deformed. Only
the latter leads to a reversible energy storage.
Therefore, the strain $\varepsilon_{ij}=u_{ij}+p_{ij}$
has two parts, the elastic and plastic one, with the
first {\em defined} as the part that changes the
energy. Hence the energy density $w(u_{ij})$ is a
function of the elastic strain $u_{ij}$, which alone
we identify as a state variable. For analogy, think of
riding a bike on a snowy path, up a steep slope. The
rotation of the wheel, containing slip and
center-of-mass motion, corresponds to the total
displacement $d$. The gravitational energy $w(d_t)$ of
the cyclist and his bike depends only on the
center-of-mass movement $d_t$, the ``elastic" or
energy-changing portion here. And the gravitational
force on the center of mass is $f_g=-\partial
w/\partial d_t$. Similarly, the elastic stress is
$\pi_{ij}=-\partial w(u_{ij})/\partial u_{ij}$. When
grains jiggle, granular deformation relax, hence
\begin{equation}\label{n1}
\partial_t u_{ij}={\rm v}_{ij}-u_{ij}/\tau,
\end{equation}
with the usual elastic term ${\rm
v}_{ij}\equiv\frac12(\nabla_iv_j+\nabla_jv_i)$, and a
relaxation term $-u_{ij}/\tau$ that accounts for
plasticity. [Note because the total strain obeys
$\partial_t\, \varepsilon_{ij}={\rm v}_{ij}$, the
evolution of the plastic strain
$p_{ij}\equiv\varepsilon_{ij}-u_{ij}$ is also fixed by
Eq~(\ref{n1}), and given as $\partial_t\,
p_{ij}=u_{ij}/\tau$.] To understand how plasticity
comes about, consider first the following scenario
with $\tau=$ constant. If a granular medium is
deformed quickly enough by an external force, leaving
little time for relaxation, $\int (u_{ij}/\tau)\,{\rm
d}t\approx0$, we have
$u_{ij}\approx\varepsilon_{ij}=\int\,{\rm v}_{ij}{\rm
d}t$ and $p_{ij}=0$ right after the deformation. The
built-up in elastic energy and stress $\pi_{ij}$ is
maximal. If released at this point, the system snaps
back toward its initial state, as prescribed by
momentum conservation, $\partial_t\,(\rho {\rm
v}_i)+\nabla_j\pi_{ij}=0$, displaying an elastic,
reversible behavior. But if the system is being held
still ($\partial_t\,\varepsilon_{ij}={\rm v}_{ij}=0$)
long enough, the elastic strain $u_{ij}$ will relax,
$\partial_tu_{ij}=-u_{ij}/\tau$, while the plastic
strain grows accordingly,
$\partial_tp_{ij}=u_{ij}/\tau$. When $u_{ij}$
vanishes, elastic energy $w(u_{ij})$ and stress
$\pi_{ij}$ are also gone, implying $\partial_t\,(\rho
{\rm v}_i)=0$. The system now stays where it is when
released, and no longer returns to its original
position. This is what we call plasticity.

However, $1/\tau$ is not a constant in sand: It grows
with the jiggling of the grains (as the deformation is
lost more quickly) and vanishes if they are at rest.
If we quantify the jiggling by the associated kinetic
energy, or (via the gas analogy) by a granular
temperature $T_g$, we could account for this by
assuming $1/\tau\sim T_g$.

As discussed above, a shear rate would jiggle the
grains, giving rise to $T_g$. For a constant rate, an
expression of the form $T_g\sim \sqrt{{\rm v}_{ij}{\rm
v}_{ij}}\equiv||{\rm v}_{s}||$ is appropriate [see
Eq~(\ref{n9}) below]. Inserting $1/\tau=\Lambda||{\rm
v}_{s}||$ (with $\Lambda$ the proportionality
coefficient) into Eq~(\ref{n1}), we obtain the
rate-independent expression, $\partial_t u_{ij}={\rm
v}_{ij}-\Lambda u_{ij}||{\rm v}_{s}||$. Being a
function of $u_{ij}$, the stress $\pi_{ij}(u_{ij})$
therefore  obeys the evolution equation,
\begin{eqnarray}\label{n2}
{\partial_t}{\pi_{k\ell}}=M_{k\ell ij}
\partial_tu_{ij}=M_{k\ell
ij}({\rm v}_{ij}-\Lambda u_{ij}||{\rm v}_{s}||),
\\\nonumber
M_{k\ell ij}\equiv{\partial\pi_{k\ell}}/ {\partial
u_{ij}}\equiv{\partial^2 w}/{\partial u_{ij}\partial
u_{k\ell}},
\end{eqnarray}
which clearly possesses the structure of
hypoplasticity~\cite{Kolym-1,Kolym-2}, a
state-of-the-art engineering model originally adopted
because sand is incrementally nonlinear, and responds
with different stress increases depending on whether
the load is being increased (${\rm
v}_{ij}>0,\,\,||{\rm v}_{s}||>0$) or decreased (${\rm
v}_{ij}<0,\,\,||{\rm v}_{s}||>0$). It is reassuring to
see that the realism of hypoplasticity is based on the
elementary physics that granular deformation is
depleted if the grains jiggle; and it is satisfying to
realize that the complexity of plastic flows derives
from the simplicity of stress relaxation.

Under cyclic loading of small amplitudes, because the
shear rate is not constant, $T_g$ oscillates and never
has time to grow to its stationary value of
$T_g\sim||{\rm v}_{s}||$. Therefore, the plastic term
$u_{ij}/\tau\sim T_g$ remains small, and the system's
behavior is rather more elastic than rendered by
Eq~(\ref{n2}).

The complete equation for $u_{ij}$ is in fact somewhat
more complex,~\footnote{even assuming hard grains,
with an elastic strain $u_{ij}$ that is typically
tiny, of order $10^{-4}$ }
\begin{eqnarray}{\rm d}_t
u_{ij}=(1-\alpha){\rm v}_{ij}-u_{ij}^*/\tau-
u_{\ell\ell}\,\delta_{ij}/\tau_1, \label{n3}\\
1/\tau=\lambda T_g,\quad 1/\tau_1=\lambda_1 T_g,
\label{n2a}\end{eqnarray}
where $u_{ij}^*$ is the deviatoric (or traceless) part of
$u_{ij}$ and ${\rm d}_t\equiv\partial_t+{\rm
v}_k\nabla_k$. The modifications are: (1)~The relaxation
time for $u_{ij}^*$ and $u_{\ell\ell}$ are different.
(2)~A shear rate ${\rm v}_{ij}$ yields an elastic
deformation rate ${\rm d}_t u_{ij}$ that is smaller by
the factor of $(1-\alpha)$.

In contrast to strain relaxation $\sim u_{ij}/\tau$ that
is irreversible, $\alpha$ accounts for reversible
processes (such as rolling). Without relaxation, elastic
and total strain are always proportional, and for say
$\alpha=2/3$, $u_{ij}$ is a third of $\varepsilon_{ij}$.
Circumstances are then reversible and quite analogous to
a solid -- aside from the fact that one needs to move
three times as far to achieve the same deformation. So
the physics accounted for by $\alpha$ is akin to that of
a lever. [This is also the reason why the stress, or
counter-force, is smaller by the same factor,
see~Eq(\ref{n12}).] Note since any granular plastic
motion such as rolling and slipping, be it reversible or
irreversible, become successively improbable when the
grains are less and less agitated, we expect
\begin{equation}
\alpha(T_g)\to0, \text{\quad for\quad} T_g\to0,
\end{equation}
implying granular media are fully elastic at vanishing
granular temperature.

\subsection{Mass, Entropy and Granular Entropy}

The energy density $w_0(s,\rho)$ of a quiescent
Newtonian fluid depend on the entropy density $s$ and
mass density $\rho$, both per unit volume. Defining
the temperature and chemical potential as
$T\equiv\partial w_0/\partial s|_\rho$ and
$\mu\equiv\partial w_0/\partial\rho|_s$, we note that
they can be computed only if the functional dependence
of $w_0(s,\rho)$ is given. The pressure, a prominent
quantity in fluid mechanics, is also a conjugate
variable, as it is given by $P\equiv\partial\bar
w/\partial v$ at constant $sv$, where  $v\equiv1/\rho$
is the specific volume, $\bar w\equiv w_0v$ the energy
per unit  mass. Again, $P$ is given once $w_0(s,\rho)$
is. (Note it is not independent from $\mu$ and $T$,
since it may be written as $P=-w_0+Ts+\mu\rho$.)

The conserved energy $w$ depends also on the momentum
density $g_i=\rho {\rm v}_i$, and is generally given
as $w=w_0+g^2/2\rho$. So the complete set of variables
is given as  $s,\rho$ and $g_i$, and the hydrodynamic
theory of Newtonian fluids consists of five evolution
equations for them. Being a structure of an actual
theory, these equations contain $w_0, P$, also $T,\mu,
{\rm v}_i\equiv\partial w/\partial g_i$. They are
closed only when $w_0$ is
specified.~\footnote{Frequently, it is enough to know
$w_0$ in a small environment around given values of
$s$ and $\rho$, or equivalently, of $T$ and $P$, if
these are taken as the independent variables.}

In continuum-mechanical theories, the entropy $s$ is
not always given the attention it deserves. The basic
facts underpinning its importance are: The conserved
energy $w$ is,  in equilibrium, equally distributed
among all degrees of freedom, macroscopic ones such as
$\rho,g_i$, and microscopic ones such as electronic
excitations or phonons (ie, short wave length sound
waves). The entropy $s$ is the macroscopic degree of
freedom that subsumes all microscopic ones (typically
of order $10^{23}$), and accounts for the energy
contained in them. Off equilibrium, energy is more
concentrated in a few degrees of freedom, typically
the macroscopic ones. The one-way, irreversible
transfer of energy from the macroscopic to the
microscopic ones -- in fluid mechanics from $\rho,g_i$
to $s$ -- is what we call dissipation, and the basic
cause for irreversibility. A proper account of
dissipation must consider the variable $s$, its
conjugate variable $T$, and the entropy production $R$
[with $R/T$ denoting the rate at which entropy is
being increased, see Eq~(\ref{n5})]. This remains so
for systems (such as granular media) that typically
execute isothermal changes.

The energy density of a solid depends on an additional
tensor variable, the elastic strain
$u_{ij}=\varepsilon_{ij}$, which in crystals is very
close to the total strain. The associated conjugate
variable $\pi_{ij}\equiv-\partial w_0/\partial u_{ij}$ is
the elastic stress -- where linear elasticity, or
$\pi_{ij}\sim u_{ij}$, represents the simplest case. The
hydrodynamic theory of solids consists of eleven
evolution equations, for the variables
$s,\rho,g_i,u_{ij}$, which in their structure contain the
conjugate variables $T,\mu,{\rm v}_i,\pi_{ij}$.

Displaying solid and liquid behavior, granular media have
the same variables -- in addition to the one that
quantifies granular jiggling, for which a scalar should
suffice if the motion is sufficiently random. We call it
{\em granular entropy} $s_g$, and define it to contain
all inter-granular degrees of freedom: the stochastic
motion of the grains (in deviation from the smooth,
macroscopic velocity) and the elastic deformation
resulting from collisions. We divide all microscopic
degrees of freedom contained in $s$ into
the~\footnote{Typical inner granular degrees of freedom
are again phonons and electronic excitations.} inner- and
inter-granular ones, $s-s_g$ and $s_g$, with the
conjugate variables $T\equiv\partial w_0/\partial
(s-s_g)$ and $T_g\equiv\partial w_0/\partial s_g$.
Equilibrium is established, when both temperatures are
equal, and $s_g$ vanishes. (There are overwhelmingly more
inner than inter granular degrees of freedom. When all
degrees have the same amount of energy, there is
practically no energy left in $s_g$.) The equilibrium
conditions are:
\begin{equation}\label{n4a}
s_g=0,\quad\bar T_g\equiv T_g-T=0.
\end{equation}
As zero is the value $s_g$ invariably returns to if
unperturbed, it is an energy minimum. Expanding the
$s_g$-dependent part of the energy $w_2\equiv
w-w(s_g=0)$, we take~\footnote{With ${\rm d}w_2=T{\rm
d}(s-s_g)+T_g{\rm d}s_g=T{\rm d}s+\bar T_g{\rm d}s_g$,
we have $T_g\equiv\partial w_2/\partial s_g|_{s-s_g}$
and $\bar T_g\equiv\partial w_2/\partial s_g|_s$.}
\begin{equation}\label{n4c}
w_2(s,\rho,s_g)=s_g^2/(2\rho b),\quad \bar
T_g\equiv\partial w_2/\partial s_g|_s=s_g/\rho b,
\end{equation}
with $b(s, \rho)>0$. So the twelve independent variables
are: $s,s_g,\rho,g_i,u_{ij}$ and the hydrodynamic theory
consists of evolution equations for them all, of which
six are given by Eq~(\ref{n3}). The rest will be given in
section~\ref{GSH}. These equations will contain $w_0$ and
the conjugate variables: $T,\bar T_g,\mu,{\rm
v}_i,\pi_{ij}$, also the pressure, given as
\begin{equation}\label{n4}
P_T\equiv-\partial\bar w_0/\partial v
\equiv-w_0+\mu\rho+sT+s_g\bar T_g,
\end{equation}
with the derivative taken at constant $sv, s_gv$ and
$u_{ij}$. As we shall see in Eq~(\ref{den11}), this is
the pressure that accounts for the contribution of
agitated grains.

\subsection{History Dependence and Fabric Anisotropy}

Finally, some remarks about the special role of the
density in granular behavior. First, it is quite
independent of the compression $u_{\ell\ell}$: Plastic
motion rearranges the packaging and change the density
by up to 20\%, without any elastic compression.
Second, the local density only changes if there is
some jiggling and agitation of the grains, $\bar
T_g\not=0$. Even when non-uniform, a given density
remains forever if the grains are at rest. So, if a
pouring procedure produces a density inhomogeneity,
this will persist as long as the system is left
unperturbed, providing an explanation for the history
dependence of static stress distribution. Sometimes,
these density inhomogeneities have a preferred
direction, say, a density gradient along $\hat x$.
With density-dependent elastic coefficients, the
system will then mimic fabric anisotropy, displaying a
stress-distribution reminiscent of an anisotropic
medium -- even when it consists of essentially round
grains and the applied stress is isotropic. Our
working hypothesis, given a preliminary validation in
section~\ref{3geo}, is that both effects are covered
by density inhomogeneities. The static stress of a
sand pile is calculated there and compared to
experiments for two densities, the first uniform and
the second with a reduced core density, which we argue
is a result of different pouring procedures, being
rain-like and funnel-fed, respectively.

\section{Granular Solid Hydrodynamics (GSH)\label{GSH}}
This section presents the remaining six evolution
equations. They will be explained but not derived,
see~\cite{JL6} for more details and the complete
derivation.
\subsection{Entropy Production} The evolution equation
for the entropy density $s$ is
\begin{eqnarray}\label{n5}
\partial_ts+\nabla_i(s{\rm v}_i-\kappa\nabla_iT)
=R/T,
\\ \label{n6}
R=\eta {\rm v}_{ij}^*{\rm v}_{ij}^*+\zeta {\rm
v}_{\ell\ell}^2+ \kappa(\nabla_iT)^2
\\\nonumber
+\gamma \bar T_g^2
+\beta(\pi^*_{ij})^2+\beta_1\pi_{\ell\ell}^2.
\end{eqnarray}
Eq~(\ref{n5}) is the balance equation for the entropy
$s$. It is (with $R$ unspecified) quite generally
valid, certainly so for Newtonian fluids and solids.
The term $sv_i$ is the convective one that accounts
for the transport of entropy with the local velocity,
and $\kappa\nabla_iT$ is the diffusive term that
becomes operative in the presence of a temperature
gradient. $R/T>0$ is the source term. It vanishes in
equilibrium, and is positive-definite off it, to
account for the fact that the conserved energy $w$
always goes from the macroscopic degrees of freedom to
the microscopic ones, $w\to s$.

The functional dependence of $R$ changes with the
system. In liquids, $R$ is fed by shear and
compressional flows, and by temperature
gradients~\cite{LL6}, as depicted by the first line of
Eq~(\ref{n6}). In equilibrium, we have ${\rm
v}_{ij},\nabla_iT=0$; off it, the quadratic form with
positive shear and compressional viscosity,
$\eta,\zeta>0$ and  heat diffusion coefficient,
$\kappa>0$, ensures that the entropy $s$ can only
increase. In fact, the terms of the first line are, in
an expansion of $R$, the lowest order positive ones
that are compatible with isotropy.

The second line of Eq~(\ref{n6}), with
$\gamma,\beta,\beta_1>0$, displays the additional
dissipative mechanisms relevant for granular media. As
discussed in the introduction, a finite $\bar T_g$ or
$\pi_{ij}$, indicating some jiggling or deformation of
the grains, will both relax and give rise to entropy
production. Since granular stress $\pi_{ij}$ will not
dissipate for $\bar T_g=0$, we require
$\beta,\beta_1\to0$ for $\bar T_g\to0$.

Being part of the total entropy, the granular entropy
$s_g$ obeys a rather similar equation, though it needs to
account for a two-step irreversibility, $w\to s_g\to s$,
the fact that the energy goes from the macroscopic
degrees of freedom to the mesoscopic, inter granular ones
of $s_g$, and from there to the microscopic, inner
granular ones of $s$, never backwards,
\begin{eqnarray}\label{n7}
\partial_ts_g+\nabla_i(s_g{\rm v}_i-\kappa_g\nabla_i\bar T_g)
=R_g/\bar T_g,\\ R_g=\eta_g {\rm v}_{ij}^*{\rm v}_{ij}^*
+\zeta_g {\rm v}_{\ell\ell}^2+ \kappa_g(\nabla_i\bar
T_g)^2-\gamma \bar T_g^2. \label{n8}
\end{eqnarray}
Eq~(\ref{n7}) has the exact same form as
Eq~(\ref{n5}), so do the first three terms of $R_g$.
But $R_g$ also has a negative contribution. The three
positive ones, with $\eta_g,\zeta_g,\kappa_g>0$,
account for $w\to s_g$, how shear and compressional
flows, and gradients in the granular temperature
produce $s_g$, the jiggling of the grains. The
negative term $-\gamma \bar T_g^2$ accounts for
$s_g\to s$, how the jiggling turns into heat. There is
the same term, though with negative sign, in $R$,
because the same amount of energy arriving at $s$ must
have left $s_g$. As emphasized, all transport
coefficients $\eta, \eta_g, \zeta, \zeta_g, \kappa,
\kappa_g, \gamma, \beta, \beta_1$ are functions of the
state variables (which may alternatively be taken as
$T, \bar T_g, \rho$, $\pi_{\ell\ell}$ and
$\pi_s^2\equiv\pi_{ij}^*\pi_{ij}^*$).

In the stationary and uniform limit, for $R_g=0$ and
$\nabla_iT_g=0$, macroscopic flows produce the same
amount of granular entropy as is leaving, implying
\begin{equation}\label{n9}
\gamma\bar T_g^2=\eta_g {\rm v}_{ij}^*{\rm
v}_{ij}^*+\zeta_g {\rm v}_{\ell\ell}^2.
\end{equation}
This is the relation employed to arrive at Eq~(\ref{n2}),
showing that hypoplasticity holds in the limit of
stationary shear rates. Given a shear rate, part of its
energy will turn into $s_g$, which in turn will leak over
to $s$. At the same time, some of the flow's energy will
heat up the system directly, with the ratio of the two
dissipative channels parameterized by $\eta/\eta_g$ and
$\zeta/\zeta_g$. In dry sand, $\eta,\zeta$ are probably
negligible und shall be neglected below -- though they
should be quite a bit larger in sand saturated with
water: A macroscopic shear flow of water implies much
stronger microscopic ones in the fluid layers between the
grains, and the dissipated energy contributes to $R$.

Finally, we consider the $\bar T_g$-dependence of
$\eta_g,\zeta_g,\gamma$. Expanding them,
\begin{equation}\label{n13}
\eta=\eta_0+\eta_1\bar T_g,\quad
\zeta_g=\zeta_0+\zeta_1\bar T_g,
\quad\gamma=\gamma_0+\gamma_1\bar T_g,
\end{equation}
we shall assume $\eta_0,\zeta_0=0$, because
\begin{itemize}
  \item $R_g$ then stays well defined for $\bar T_g\to0$,
  see Eq~(\ref{n8});
  \item Viscosities typically vanish with
temperature;
  \item This fits the Bagnold scaling;
  \item For
$\gamma_0\gg\gamma_1\bar T_g$ and
$\gamma_0\ll\gamma_1\bar T_g$, respectively, we have
from Eq~(\ref{n9}), for ${\rm v}_{\ell\ell}=0$,
\begin{equation}\label{den17}
\bar T_g=(\eta_1/\gamma_0)\,|{\rm v}_{ij}^*|^2,\quad \bar
T_g=\sqrt{\eta_1/\gamma_1}\,|{\rm v}_{ij}^*|.
\end{equation}
This ensures the existence of an elastic regime at
vanishing $\bar T_g$, see section~\ref{hypo}.

  \end{itemize}

\subsection{Conservation Laws} The three evolution
equations left to be specified are conservation laws, for
mass, energy and momentum,
\begin{eqnarray}\label{n10}
\partial_t\rho+\nabla_i(\rho {\rm v}_i)=0,\quad
\partial_t w+\nabla_iQ_i=-\rho v_i\nabla_i\phi,\\
\partial_t(\rho {\rm v}_i)+\nabla_i(\sigma_{ij}+\rho
{\rm v}_iv_j)=-\rho\nabla_i\phi, \label{n11}
\end{eqnarray}
where $\phi$ is the gravitational potential (on the
earth surface, we have $-\nabla_i\phi=G_i$, the
gravitational constant pointing downwards). Without
specifying the fluxes $Q_i,\sigma_{ij}$, these
equations are always valid, quite independent of the
system, and express the simple fact that being locally
conserved quantities (in the absence of gravitation),
energy, momentum and mass obey continuity equations.
The basic idea of the hydrodynamic theory is to
require the structure of the fluxes $Q_i, \sigma_{ij}$
to be such that, with the temporal derivatives of the
variables given by
Eqs~(\ref{n3},\ref{n5},\ref{n7},\ref{n10},\ref{n11}),
the thermodynamic relation
\begin{eqnarray*}
\partial_t w(s,s_g,\rho,g_i,u_{ij})=(\partial
w/\partial s)\partial_t s+(\partial w/\partial
s_g)\partial_t s_g+(\partial w/\partial\rho)\partial_t
\rho\\+(\partial w/\partial g_i)\partial_t
g_i+(\partial w/\partial u_{ij})\partial_t
u_{ij}\\=T\partial_t s+\bar T_g\partial_t
s_g+\mu\partial_t \rho+v_i\partial_t
g_i-\pi_{ij}\partial_t u_{ij}\quad\,\,
\end{eqnarray*}
is identically satisfied, irrespective of $w$'s
functional form. This is a rather confining bit of
information, enough to uniquely fix the two fluxes as
\begin{eqnarray}\label{n10a}
Q_i=(w+P_T){\rm v}_i+\sigma_{ij}{\rm v}_j -\kappa
T\nabla_iT-\kappa_g\bar T_g \nabla_i\bar T_g,
\\ \label{n12}
\sigma_{ij}=(1-\alpha)\pi_{ij}+(P_T-\zeta_g {\rm
v}_{\ell\ell})\delta_{ij}-\eta_g {\rm v}_{ij}^*,
\end{eqnarray}
with $P_T$ given by Eq~(\ref{n4}), and ${\rm v}_{ij}^*$
being the deviatory (or traceless) part of  ${\rm
v}_{ij}$. (For details of derivation see~\cite{JL6}.)
Although now specified to fit granular physics as
codified in Eqs~(\ref{n3},\ref{n5},\ref{n7}), these are
still fairly general results, valid irrespective what
concrete form $w$ assumes. Moreover, they also nicely
demonstrate the dependence on the number and types of
variables: Eliminating $s_g$, or equivalently, taking
$\bar T_g=0$, in Eqs~(\ref{n4},\ref{n10a},\ref{n12}), one
obtains the solid hydrodynamics.~\footnote{In solids,
density change and compression are not usually
independent. We may account for this by formally setting
$P_T=0$.} Further eliminating $u_{ij}$ by taking
$\pi_{ij}=0$ leads to the fluid hydrodynamics.

Focusing on the plastic motion, the standard approach
(especially the thermodynamic consideration by Houlsby
and coworkers,~\cite{Houlsby}) employs the plastic
strain $p_{ij}\equiv\varepsilon_{ij}-u_{ij}$ as the
independent variable. Although this starts from the
same insight about plastic motion, the connection
between elastic strain, stress and energy, so similar
in solids and granular media, with formulas that hold
for both systems, is lost -- or at least too well
hidden to be useful, see also the discussion in
section~\ref{Houl}.

Enforcing a velocity gradient ${\rm v}_{ij}$, the rate
of work being received by the system is
$-\sigma_{ij}{\rm v}_{ij}=-[(1-\alpha)\pi_{ij}
+P_T\delta_{ij}]{\rm v}_{ij}+[\zeta_g {\rm
v}_{\ell\ell}{\rm v}_{\ell\ell}
+\eta_gv^*_{ij}v^*_{ij}]$, see Eq~(\ref{n10a}). Of
these, the terms in the first square brackets, being
proportional to the velocity and hence odd under time
inversion, are reactive; while those $\sim{\rm v}^2$
in the second bracket are even and dissipative. Work
received via an odd term will leave if its sign is
changed by inverting time's direction; work received
via an even term stays, as happens only with
dissipative processes. The reappearance of the same
factor $(1-\alpha)$ as in Eq~(\ref{n3}) is not an
accident, but required by energy conservation. If the
same velocity leads to an elastic deformation that is
smaller by $(1-\alpha)$, then just as with a lever,
the force counteracting this deformation
$\sigma_{ij}=(1-\alpha)\pi_{ij}+\cdots$ is smaller by
the same factor.

This concludes the derivation of the structure of {\sc
gsh}, or granular solid hydrodynamics, given by
Eqs~(\ref{n3},\ref{n4}),
(\ref{n5},\ref{n6},\ref{n7},\ref{n8}) and
(\ref{n10},\ref{n11},\ref{n10a},\ref{n12}).

\section{Validation of {GSH}}
The advantage of {\sc gsh} is two-fold, its clear
connection to the elementary granular physics as spelt
out in the introduction, and more importantly, the
stringency of its structure. It cannot be changed at
will to fit experiments, without running into
difficulties with general principles. The only
remaining liberty is the choice of the functional
dependence for the energy and some transport
coefficients. As this implies much less wiggle room
than with typical continuum-mechanical models, any
agreement with experimental data is less designed,
``hand-crafted," and more convincing, especially with
respect to the starting physics.

In what follows, we shall fist examine granular
statics, for a medium at rest, $T_g=0$, then go on to
granular dynamics, with enforced flows or stress
changes, and some accompanying jiggling, $T_g\not=0$.
An expression for the conserved energy $w$ will be
proposed that, in spite of its relative simplicity,
reproduces many important granular features when
embedded into {\sc gsh}. As discussed above
Eq~(\ref{n4a}), we divide $w$ into three parts: the
micro-, macro- and mesoscopic ones,
\begin{equation}\label{den3}
w=w_0(s,\rho) +[w_1(u_{ij},\rho,g_i)+ g^2/2\rho]
+w_2(s_g,\rho).
\end{equation}
The first~\footnote{Assuming that only $w_0$ depends
on $s$ neglects effects such as thermal expansion
which, however, can be easily included if needed.}
accounts for the inner-granular degrees of freedom,
all subsumed as heat into the true entropy $s$. We
take $w_0=\langle E(s)/m\rangle\rho$, where $E(s)$ is
the energy of a grain, $m$ its mass, and $\langle
\rangle$ denotes the average.  The second consists of
the contributions from the macroscopic variables of
momentum density $g_i$ and the elastic strain
$u_{ij}$, where $w_1$ is given by Eq~(\ref{den2})
below. The third, $w_2(s_g,\rho)$ of Eq~(\ref{n4c}),
is further specified in section~\ref{f_2}. It accounts
for the inter-granular degrees of freedom, the
mesoscaled, strongly fluctuating elastic and kinetic
contributions.

\subsection{Granular Statics, $\bf{T_g=T}$} Given
an energy $w_1(u_{ij})$, we can use the stress
$\pi_{ij}(u_{ij})\equiv-\partial w_1/\partial u_{ij}$
and $u_{ij}=\frac12 (\nabla_iU_j+\nabla_j U_i)$ to
close the stress balance $\nabla_j\pi_{ij}(r_i)=\rho
G_i$, and determine $\pi_{ij}(r_i)$ with appropriate
boundary conditions. As this is done without any
knowledge of the plastic strain, we may with some
justification call this {\em granular
elasticity}~\cite{ge-1}.

The relation $u_{ij}=\frac12(\nabla_iU_j+ \nabla_jU_i)$
remains valid because of the following reasons: In an
elastic medium, the stressed state is characterized by a
displacement field from a unique reference state, in
which the elastic energy vanishes. Because there is no
plastic deformation $U_i^p$, the total displacement is
equal to the elastic one. Circumstances appear at first
quite different in granular media. Starting from a
reference state, a stressed one is produced by the
displacement $U_i+U_i^p$, with typically $U_i^p\gg U_i$.
Due to sliding and rolling, $U_i^p$ is highly
discontinuous, but $U_i$ remains slowly varying, because
the cost in elastic energy would otherwise be
prohibitive. Fortunately, $U_i^p$ is quite irrelevant: We
have innumerable reference states, all with vanishing
elastic energy and connected to one another by purely
plastic deformations. As a result, we can, for any given
displacement $U_i+U_i^p$, switch to the reference state
that is separated from the original one by $U_i^p$, and
to the stressed one by $U_i$. Now, the circumstances are
completely analogous to that of an elastic medium.

\subsubsection{Yield Surfaces}

An important aspect of granular behavior, in the space
spanned by the variables,  is the existence of yield
surfaces. We take them to be the divide between two
regions, one in which stable elastic solutions are
possible, the other in which they are not -- so the
system must flow and cannot come to rest. A natural
and efficient way to account for yield is to code it
into the energy, a scalar. Given the stress balance,
the energy is extremal~\cite{ge-1} -- minimal if
convex and maximal if concave. Having the energy being
convex within the yield surface, and  concave beyond
it, any elastic solution that is stable within the
surface, will be eager to get rid of the excess energy
and become unstable against infinitesimal
perturbations beyond it.

\subsubsection{The Elastic Energy $\bf{w_1}$\label{w_1}}

Our present choice for the elastic energy
is~\cite{JL6,J-L-1,J-L-2,J-L-3},
\begin{equation}\label{den2}
w_1(\rho,u_{ij})={\mathcal B}\sqrt{\Delta
}\left(2\Delta^2/5+ {u_s^2}/\xi \right),
\end{equation}
where $\Delta\equiv -u_{\ell\ell}$, $u_s^2\equiv
u^*_{ij}u^*_{ij}$. The energy $w_1$ is convex only for
$u_s/\Delta\le\sqrt{2\xi}$, or equivalently $\pi
_s/P_\Delta\le\sqrt{2/\xi}$ (where
$P_\Delta\equiv\frac13\pi _{\ell \ell }$,
$\pi_s^2\equiv \pi^*_{ij}\pi^*_{ij}$), which coincides
with the Drucker-Prager condition.~\footnote{Only if
an energy expression depends on the third strain
invariant, could it possibly contain an instability at
the true Coulomb condition.} Taking $\xi=5/3$ gives a
friction angle of about $28^\circ$. We further take
${\cal B}={\cal B}_0\,{\cal B}_1(\rho)\,{\cal
C}(\rho,u_{ij})$, where ${\cal B}_0$ is a constant,
and
\begin{eqnarray}\label{den4}{\cal B}_1&=&
\left[(\rho -\rho^*_{\ell p})/(\rho _{cp}-\rho)\right]
^{0.15},
\\\label{den5}
2{\cal C}&=&1+\tanh [(\Delta _0-\Delta)/\Delta _1].
\end{eqnarray}
The coefficient ${\cal B}_1$ diverges for the ``random
closed-pack" density, $\rho_{cp}$, and is convex only
between $\rho_{cp}$ and the ``random loose pack"
density $\rho_{\ell p}$. [$\rho^*_{\ell p}$ is a
constant chosen to yield the right value for
$\rho_{\ell p}$ with the relation $\rho_{\ell p}\equiv
(11\rho_{cp}+9\rho^*_{\ell p})/20$.] It accounts for
(1)~the lack of elastic solutions for $\rho<\rho_{\ell
p}$, when the grains loose contact with one another;
(2)~the stiffening of granular elasticity with growing
density, until it (as an approximation for becoming
very large) diverges at $\rho_{cp}$.

\begin{figure}[t]
  \includegraphics[scale=.7]{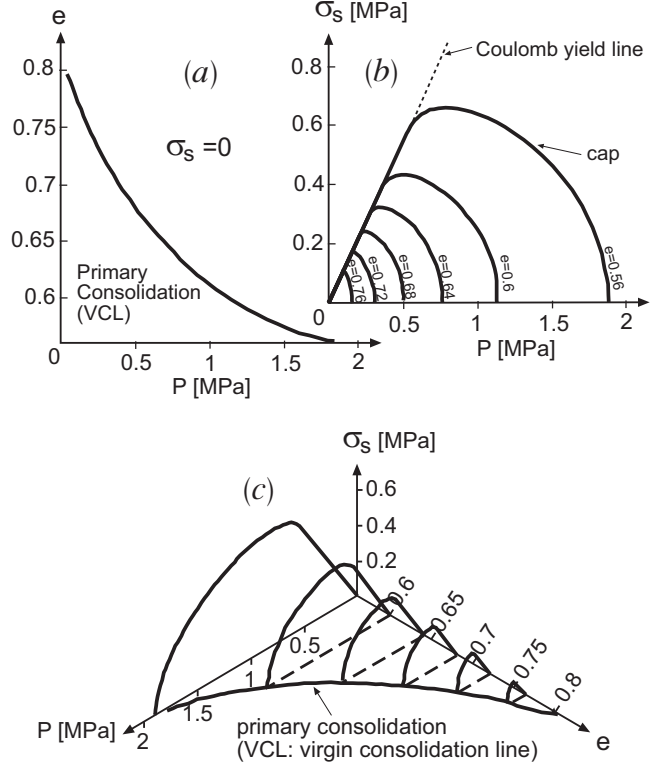}
  \caption{\label{fig1}Yield surfaces as coded
in the energy of Eqs~(\ref{den2},\ref{den4},\ref{den5}),
a function of the pressure, shear stress, and void ratio.
(a):~The {\em virgin consolidation line}. (b):~The
bending of the Coulomb yield line, as a function of $e$.
(c):~Combination of (a) and (b).}
\end{figure}

With $\Delta_0, k_1, k_2, k_3$ being constants, and
$\Delta _0=k_1\rho -k_2u_s^2-k_3$, we have ${\cal
C}=1$ for $\Delta\ll\Delta_0$, and ${\cal C}=0$ for
$\Delta\gg\Delta_0$. It changes from 1 to 0 in a
neighborhood of $\Delta_1$ around $\Delta_0$,
destroying the energy's convexity there. Taking
$\Delta_0$ to grow with the density and fall with
$u_s^2$ limits the region of stable elastic solutions
to sufficiently small $\Delta$-values, reproducing the
virgin consolidation curve and the so-called caps at
varying void ratios $e$, see Fig~\ref{fig1}.

\subsubsection{Stress Distribution for Silos,
Sand Piles and Point Loads\label{3geo}}

Three classic cases, a silo, a sand pile and a granular
sheet under a point load, are solved employing the stress
expression derived from the energy of  Eq~(\ref{den2}),
producing rather satisfactory agreement with experiments.

{\bf Silos\quad} For tall silos, the classic approach is
given by Janssen, who starts from the assumption that the
ratio between the horizontal and vertical stress is
constant, $k_J=\sigma_{rr}/\sigma_{zz}$. Assuming in
addition that $\sigma_{zz}$ only depends on $z$, not on
$r$, Janssen finds the vertical stress $\sigma_{zz}$
saturating exponentially with height -- a result well
verified by observation. (He leaves $\sigma_{rz}$ and all
three radial components: $\sigma_{\theta\theta},
\sigma_{r\theta}$ and $\sigma_{z\theta}$ undetermined.)
Having calculated $\sigma_{zz}$, one needs the value of
$k_J$ to obtain $\sigma_{rr}$, usually provided by
$k_J\simeq 1-\sin \varphi$, with $\varphi$ the friction
angle measured in triaxial tests. This makes $\varphi$
the only bulk material parameter in silo stress
distributions. We shall refer to this as the Jaky
formula, although it is also attributed to K\'{e}zdi. Being
important for the structural stability of silos, this
formula is (with a safety factor of 1.2) part of the
construction industry standard, see eg. DIN 1055-6, 1987.
We believe this formula goes well beyond its practical
relevance, that it is a key to understanding granular
stresses, because it demonstrates the intimate connection
between stress distribution and yield, a connection that
has not gained the wide attention it deserves. Starting
from Eq~(\ref{den2}), we calculated~\cite{ge-2} all six
components of the stress tensor, verifying the Janssen
assumptions to within 1\%, and found the Janssen constant
$k_J$ well rendered by the Jaky formula.

{\bf Point Loads\quad} The stress distribution at the
bottom of a granular layer exposed to a point force at
its top is calculated~\cite{ge-2} employing
Eq~(\ref{den2}), without any fit parameter. Both vertical
and oblique point forces were considered, and the results
agree well with simulations and experiments using
rain-like preparation. In addition, the stress
distribution of a sheared granular layer exposed to the
same point force is calculated and again found in
agreement with experimental data, see~\cite{ge-2} for
more details and references.

{\bf Sand Piles\quad} The fact that the pressure
distribution below sand piles and wedges, instead of
always displaying a single central peak, may sometimes
show a dip, has intrigued and fascinated many physicists,
prodding them to think more carefully and deeply about
sand. Recent experimental investigations established the
following connection: A single peak results when the pile
is formed by rain-like pouring from a fixed height; the
dip appears when the pile is formed by funneling the
grains onto the peak, from a shifting funnel always
hovering slightly above the
peak. 
Employing Eq~(\ref{den2}) to consider the stress
distribution in sand wedges, we found the pressure at
the bottom of the pile to show a single central peak
if a uniform density is assumed. The peak turns into a
pressure dip, if density inhomogeneity, with the
center being less compact, is assumed. The two
calculated pressure distributions are remarkably
similar to the measured ones, see~\cite{ge-1}. The
nonuniform density, we believe, is a consequence of
pile formation using the hovering funnel: Since the
funnel is always just above the peak, the grains are
placed there with very little kinetic energy,
resulting in a center region below the peak that has a
low density. Those grains that do not find a stable
position roll down the slope and gather kinetic
energy. When they crash to a stop at the flanks, they
compact the surrounding, achieving a much higher
density.

\subsection{Granular Dynamics, $\bf{T_g\not=T}$}

If a granular medium is exposed either to stress
changes, or a moving boundary, the grains will flow,
displaying both a smooth, macroscopic velocity, ${\rm
v}_i\not=0$, and some stochastic jiggling,
$s_g\sim\bar T_g\not=0$. Then the following effects
will come into play: First, the energy is extended by
a $s_g$-dependent contribution, $w_2(s_g,\rho)$, see
Eq~(\ref{n4c}). Second, the transport coefficients
of~Eq~(\ref{n13}) become finite. Most importantly,
third, the relaxation times $\tau,\tau_1$ of
Eq~(\ref{n3}) are no longer infinite, implying the
presence of plastic flows.

\subsubsection{The $\bf{s_g}$-Dependent Part of the Energy\label{f_2}}
Specifying the expansion coefficient $b(\rho)$ of
Eq~(\ref{n4c}) as $b=b_0(1-\rho/\rho_{cp})^{a}$, we
find
\begin{equation}
\label{den11} P_T={a\, \rho \,b_0 \bar T_g^2}
{(1-\rho/\rho_{cp})^{a-1}} ({\rho}/{2\rho_{cp}})
\end{equation}
by employing Eq~(\ref{n4}). The density dependence of
the expansion coefficient $b(\rho)$ is chosen such
that it reproduces the observed volume-dilating
pressure contribution $P_T\sim f_2/(\rho_{cp}-\rho)$
from agitated grains~\cite{Lub-1,Lub-2,Lub-3}.
However, we cannot take $a=0$ as it would imply a
diverging granular entropy $s_g$ for
$\rho\to\rho_{cp}$. Therefore, we take $a$ to be
positiv but small, where $a\approx0.1$ appears
appropriate. (Note that  with $w_0/\rho$ independent
of $\rho$ and $w_1/\rho\sim \Delta^{2.5}$ -- where
$\Delta$ rarely exceeds $10^{-4}$ -- the respective
density derivative and pressure contribution is zero
and negligibly small.)

\subsubsection{The Hypoplastic Regime\label{hypo}}
\begin{figure}[b]
\begin{center}
\includegraphics[scale=0.85]{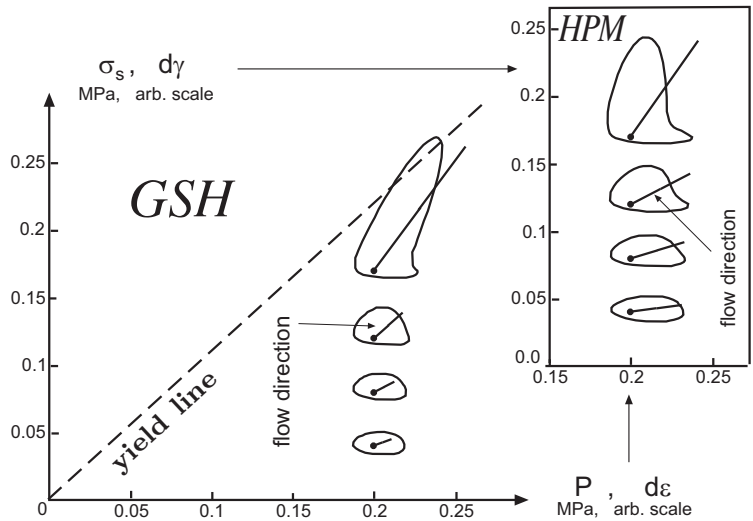}
\end{center}
\caption{The change in strain ${\rm d}\gamma\equiv({\rm
v}_{11}-{\rm v}_{33}){\rm d}t, {\rm
d}\varepsilon\equiv-(2{\rm v}_{11}+{\rm v}_{33}){\rm d}t$
for given stress rate starting from different points in
the stress space, spanned by $\sigma_s,P$, as calculated
employing (1)~{\sc gsm}, the present theory (taking
$1-\alpha=0.22$, $\tau/{\tau_1}=0.09$, ${\zeta _g}/{\eta
_g}=0.33$, $\lambda \sqrt{{\eta_g}/{\gamma}}=114$), and
(2)~{\sc hpm}, a typical hypoplastic model,
see~\cite{JL3} for more figures and details.}
\label{fig1a}
\end{figure}
We may choose our parameters such that  $\bar T_g$ is
small at typical velocities of elasto-plastic
deformations, though large enough to cover both limits
of Eq~(\ref{den17}). Then the first term of
Eq~(\ref{n12}) dominates, because all other terms
($\sim P_T,\eta_g,\zeta_g$) are of order $\bar T_g^2$.
Then we have
$\partial_t\sigma_{ij}=(1-\alpha)\partial_t\pi_{ij}
=(1-\alpha)M_{ijk\ell}\partial_t u_{ij}$, with
$\partial_t u_{ij}$ given by Eq~(\ref{n3}). Stress
relaxation, the culprit producing irreversible
plasticity, is a term $\sim \bar T_g$. For very slow
shear flows and $\bar T_g\sim||{\rm v}_{ij}||^2$
[first of Eq~(\ref{den17})], it is quadratically small
and negligible. This is the elastic regime. At
somewhat faster shear flows, the relation $\bar
T_g\sim||{\rm v}_{ij}||$ [second of Eq~(\ref{den17})]
renders $\partial_t\sigma_{ij}$ rate-independent,
giving it the basic structure of hypoplasticity,
Eq~(\ref{n3}). Comparing this results to a
state-of-the-art hypoplastic model, we found
impressively  quantitative agreement, see
Fig~\ref{fig1a}. This is remarkable, because the
anisotropy of these figures, determined essentially by
$M_{ijk\ell}$, is a calculated quantity:
$M_{ijk\ell}\equiv\partial^2 w_1/\partial
u_{ij}\partial u_{k\ell}$, with $w_1$ given by
Eq~(\ref{den2}).

\subsubsection{The Butterfly Cycle}
\begin{figure}[t]
\includegraphics[scale=2.5]{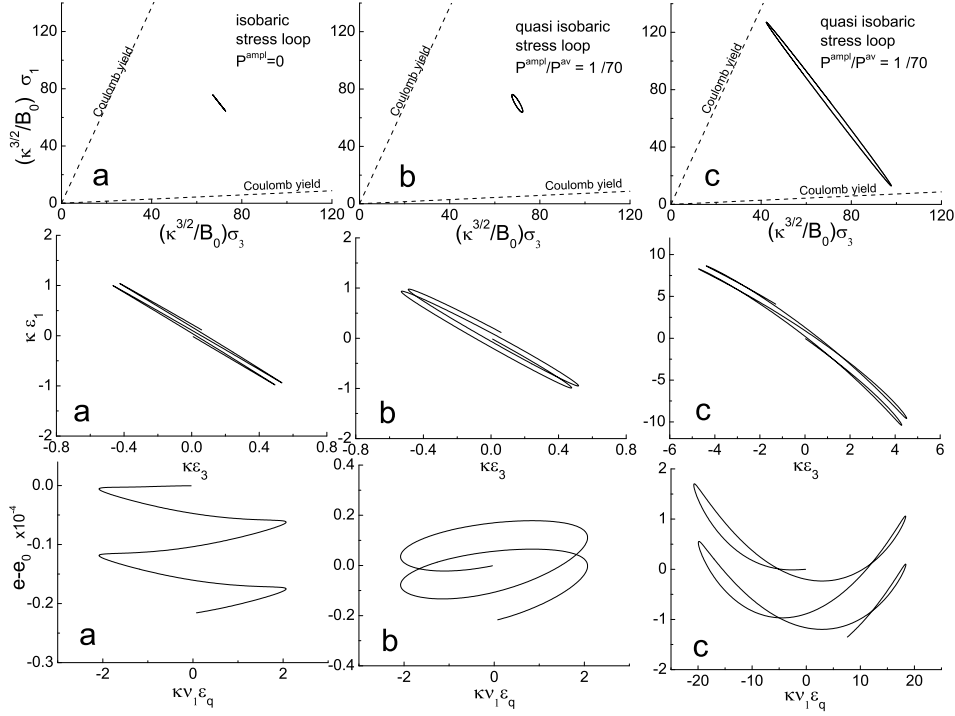}
\caption{Upper row: radial stress $\sigma _{1}$ versus
axial stress $\sigma _{3}$, rescaled by $B_{0}\kappa
^{-3/2}$. Middle row: radial strain $\varepsilon
_{1}=\int {\rm v}_{xx}dt$ versus axial strain
$\varepsilon _{3}=\int {\rm v}_{zz}dt$. Lower row:
$e-e_0$ (with $e_0$ the initial void ratio) versus shear
strain $\varepsilon_{q}=\int({\rm v}_{zz}-{\rm
v}_{xx})dt$, rescaled by $\nu _{1}\kappa $. The stress
loads are isobaric for (a) and quasi-isobaric for (b,c);
the cyclic amplitude is small for (a,b) and large for
(c). The associated strain loci and void ratio are:
sawtooth-like for (a),  coil-like for (b), butterfly-like
(or double-looped) for (c). [The large-amplitude,
isobaric plot is quite similar to (c).] }\label{Fig2}
\end{figure}

Our last example for validation is not a direct
comparison of \textsc{ghd} to some experimental data,
but rather an examination of what \textsc{ghd} does,
unforced and uncrafted, under typical elasto-plastic
deformations. It is solved numerically for stress
paths in the triaxial geometry (ie. $\sigma
_{xx}=\sigma _{yy}$, $\sigma _{ij}=0$ for $i\neq j$,
similarly for $u_{ij}$), including all energy terms
given above, except $\mathcal{C}$ of Eq~(\ref{den5})
that is set to 1 (assuming the yield surface is
sufficiently far away). All transport coefficients
depend on $T_g$ as specified, but are otherwise
constant, independent of stress and density. Also, all
variables are taken to be spatially uniform, reducing
a set of partial differential equations to ordinary
ones in time. In spite of these major simplifications,
the results as rendered in Fig~3 display such uncanny
realism that it seems obvious {\sc gsh} has captured
some important elements of granular physics.
%
%
We consider a test with the stress given as
\begin{equation}
P =P^{\text{av}}+P^{\text{ampl}}\cos \left( 2\pi
ft\right) ,  \label{5.3g} \quad q =q^{\text{ampl}}\cos
\left( 2\pi ft+\varphi \right) . 
\end{equation}
Numerical solutions were computed for isobaric test with
$P^{\text{ampl}}=0$ (ie. $P=$ constant) and
quasi-isobaric test, with
$P^{\text{ampl}}<<P^{\text{av}}$ (ie. $P\approx$
constant). The results are shown in Fig~\ref{Fig2}. they
are obtained using the dimensionless parameters: $\kappa
\equiv \sqrt{\zeta _{1}\gamma _{1}}/\rho b=18257$,
$(\gamma _{0}/\gamma _{1})^{2}$ $(\rho _{cp}b^{2}\kappa
^{3/2}/\mathcal{B}_{0}b_{0})=1.07\times 10^{-6}$,
$\lambda _{1}/\lambda =0.09 $, $\nu _{1}^{2}/2$ $\equiv
\eta _{1}/3\zeta _{1}=1$, $\lambda \sqrt{\eta _{1}/\gamma
_{1}}=114$, $\alpha =0$. The initial conditions are:
$e_{0}=0.68085$ (or $\rho _{0}=0.94\rho _{cp}$),
$v_{ij}$, $\bar T_{g}$, $\partial _{t}\bar T_{g}$,
$\partial _{t}\rho $, $\partial _{t}u_{ij}=0$. The
averaged pressure $P^{\text{av}}\equiv \sigma _{ii}/3$ is
$P^{\text{av}}=70\mathcal{B} _{0}\kappa ^{-3/2}$, and the
amplitude $q^{\text{ampl}}\equiv \sigma
_{3}-\sigma _{1}$ is $5\nu _{1}\kappa ^{3/2}q^{\text{ampl}}/6\mathcal{B}%
_{0}=10$ for (a,b) and $100$ for (c,d). The frequency of
$P^{\text{ampl}},q^{\text{ampl}}$ is $f=12(\gamma
_{0}/b\rho )$, and the phase lag between them is $\varphi
=58^{\circ }$.

\subsection{Competing Concepts and Misconceptions\label{Diff2}}
Finally, we revisit two previous approaches to come to
terms with granular behavior, {\em granular
thermodynamics} by Houlsby et al~\cite{Houlsby}, and
{\em granular statistical mechanics} by Edwards et
al~\cite{Edw}. We shall compare {\sc gsh} to both
assuming at most superficial familiarity with them.
Also, we refute some misconceptions that have become
unfortunately widespread, especially the one about
energy not being conserved in sand~[{\em sic}]. These
are at best a nuisance in exchanges with referees; and
at worst actual obstacles in the progress of our
coming to grips with granular modeling.

\subsubsection{Granular Thermodynamics\label{Houl}}
Although considerable work and thoughts have gone into
applying thermodynamics to granular media and plastic
flow, especially from Houlsby and
Collins~\cite{Houlsby}, its basic points are clear and
easy to grasp. Taking the entropy production as
\begin{equation}\label{c1}
R=\pi_{ij}\partial_t p_{ij}
\end{equation}
(where $p_{ij}$ denotes, as before, the plastic
strain), it is obvious that the usual linear Onsager
force-flux relation, $\partial_t p_{ij}\sim\pi_{ij}$,
hence $R\sim\pi_{ij}^2$, does not give a
rate-independent $R$. Therefore, Houlsby, Collins and
coworkers consider instead
\[R=\sqrt{\chi_{ijk\ell}\partial_t p_{ij}\partial_t
p_{k\ell}}=(\chi_{ijk\ell}\partial_t p_{ij}\partial_t
p_{k\ell})/{\textstyle\sqrt{\chi_{ijk\ell}\partial_t
p_{ij}\partial_t p_{k\ell}}},\]
a rate-independent expression. Equating it to
Eq~(\ref{c1}), with $\pi_{ij}=-\partial F/\partial
p_{ij}$, and $F$ being the free energy density, one
then solves for the plastic strain $p_{ij}$ with a
given $F$. One example gives $\partial_t p_{ij}\neq0$
on a yield surface, characterized by some components
of $\pi_{ij}$ being constant, and $\partial_t
p_{ij}=0$ off it.

{\sc gsh} starts with the same $R$, but possesses the
additional variable $T_g$, for which $\bar
T_g\sim||{\rm v}_s||$ frequently holds, see
Eq~(\ref{n9}). The linear Onsager force-flux relation
\begin{equation}\label{c2}
\partial_t p_{ij}=\beta\pi_{ij}\quad\text{with}
\quad \beta\sim\bar T_g,
\end{equation}
therefore suffices to yield an rate-independent $R\sim
\bar T_g\pi_{ij}^2$. Note Eq~(\ref{c2}) leads directly to
the relaxation term: Because $\partial_t
u_{ij}+\partial_t p_{ij}={\rm v}_{ij}$, we have
$\partial_t u_{ij}-{\rm
v}_{ij}=-\beta\pi_{ij}=-u_{ij}/\tau$, with $1/\tau\sim
\bar T_g$. (The last equal sign holds because
$\pi_{ij},\beta,\tau$ are all functions of $u_{ij}$, with
$\beta,\tau$ as yet unspecified.)

Summarizing, without the variable $\bar T_g$, Houlsby
and Collins needed to go beyond the well-verified and
-substantiated procedure of linear Onsager force-flux
relation to maintain rate-independence, obtaining a
plastic flow that is confined to the yield surface. In
{\sc gsh}, rate-independence arises naturally within
the confines of linear Onsager relation, producing a
plastic flow that is as realistic as hypoplasticity,
and finite also off the yield surface.

\subsubsection{Granular Statistical
Mechanics}

Generally speaking, it is important to remember that of
all microscopic degrees of freedom, the inner-granular
ones are many orders of magnitude more numerous than the
inter-granular ones. It is the former that dominate the
entropy and any entropic considerations. When revisiting
granular statistical mechanics, especially the Edwards
entropy, it is useful to keep this in mind.

Taking the entropy $S(E,V)$ as a function of the
energy $E$ and volume $V$, or ${\rm d}S=(1/T){\rm
d}E+(P/T){\rm d}V$, the authors of~\cite{Edw} argue
that {\it a mechanically stable agglomerate of
infinitely rigid grains at rest} has, irrespective of
its volume, vanishing energy, $E\equiv0$, ${\rm
d}E=0$. The physics is clear: However we arrange these
rigid grains that neither attract nor repel each
other, the energy remains zero. Therefore, ${\rm
d}S=(P/T){\rm d}V$, or ${\rm d}V=(T/P){\rm d}S\equiv
X{\rm d}S$. The entropy $S$ is obtained by counting
the number of possibilities to package grains for a
given volume, and taking it to be $e^S$. Because a
stable agglomerate is stuck in one single
configuration, some tapping or similar disturbances
are needed to enable the system to explore the phase
space.

In {\sc gsh}, the present theory, grains are neither
infinitely rigid, nor always at rest, hence the energy
contains both an elastic and a $s_g$-dependent
contribution.~\footnote{That grains neither attract
nor repel each other is accounted for by the stress
vanishing if $s_g$ and $u_{ij}$ do. Then $w_1,w_2=0$
and $w_0\sim\rho$, see Eq~(\ref{den3}), implying
$\sigma_{ij}=\partial (w_0/\rho)/\partial
(1/\rho)\delta_{ij}=0$.} And the question is whether
granular statistical mechanics is a legitimate limit
of {\sc gsh}. We are not sure, but a yes answer seems
unlikely, as both are conceptually at odds in two
points, the first more direct, the second quite
fundamental: (1)~Because of the Hertz-like contact
between grains, very little material is being deformed
at first, with the compressibility diverging at
vanishing compression. This is a geometric fact
independent of how rigid the bulk material is.
Infinite rigidity is therefore not a realistic limit
for sand. (2)~As emphasized, the number of
possibilities to arrange grains for a given volume is
vastly overwhelmed by the much more numerous
configurations of the inner granular degrees of
freedom, especially phonons. Maximal entropy $S$ for
given energy therefore realistically implies minimal
macroscopic energy, such that a maximally possible
amount of energy is in $S$ (or heat), equally
distributed among the inner-granular degrees of
freedom. Maximal number of possibilities to package
grains for a given volume is a very different
criterion.

\subsubsection{Energy Conservation}

Stemming ultimately from a loose vocabulary, some alleged
difficulties to model sand are based on fallacies that
need to be refuted here.

The essential difference between granular gas and ideal
(atomic or molecular) gas is that the particles of the
first undergo non-elastic, dissipative collisions. As a
result, their kinetic energy is not conserved, and the
velocity distribution typically lacks the time to arrive
at the equilibrium Gaussian form. Quantifying the kinetic
energy as a granular temperature $T_g$, it is therefore
hardly surprising that the fluctuation-dissipation
theorem ({\sc fdt}), formulated in terms of $T_g$, is
frequently violated. These are sound results, obtained
from a healthy but truncated model that takes the grains
as the basic microscopic entity with no heat content.
However, some of the further conclusions are deduced
forgetting this simplification, rendering them patently
absurd. These, and their {\em [refutation in italic]},
are listed below:
\begin{itemize}
  \item
As the energy is not conserved in sand, neither
thermodynamics nor the hydrodynamic method are valid.
{\em [Only the kinetic energy dissipates in granular
media, not the total energy. The latter, including
kinetic, elastic and heat contributions, remains
conserved -- as it is in any other system. And only the
conservation of total energy is important for thermo- and
hydrodynamics.]}
  \item
{\sc fdt}, along with other general principles either
derived from it or in its conceptual vicinity  (such as
the Onsager reciprocity relation) are all violated. {\em
[There are two versions of {\sc fdt}, only the one given
in terms of $T_g$ is violated, not the one in terms of
the true temperature $T$. The latter is a general
principle and always valid. For instance, the volume
fluctuation is given as $\langle\Delta
V^2\rangle=T(\partial^2 F/\partial V^2)^{-1}$, with $F$
the associated free energy, for a copper block, a single
grain, and a collection of grains.  If the grains in the
collection are jiggling, there is an extra contribution
$\sim T_g^2$ in $F$, see Eq~(\ref{n4c}), that
considerably increases the value of $\langle\Delta
V^2\rangle$. The Onsager relation remains valid because
the true {\sc fdt} holds.]}
  \item The Onsager relation is also violated
because the microscopic dynamics, the collision of the
grains, is dissipative and hence irreversible. {\em [The
true microscopic dynamics is that in terms of atoms and
molecules, the building blocks of the grains. Their
dynamics is, as in any other system, reversible.]}
\end{itemize}


\begin{thebibliography}{99}

\bibitem{Kolym-1} D. Kolymbas,
\textit{Introduction to Hypoplasticity}, (Balkema,
Rotterdam, 2000).
\bibitem{Kolym-2} D. Kolymbas, also W. Wu and D.
Kolymbas, in \textit{Constitutive Modelling of Granular
Materials} ed D. Kolymbas, (Springer, Berlin, 2000), and
references therein.

\bibitem{LL6}  L. D. Landau and E. M. Lifshitz, {\it Fluid
Mechanics} (Butterworth-Heinemann, Oxford, 1987) and
\textit{Theory of Elasticity}  (Butterworth-Heinemann,
Oxford, 1986)

\bibitem{Khal} I.M. Khalatnikov, {\em Introduction to
the Theory of Superfuidity}, (Benjamin, New York 1965).
\bibitem{hydro-1}S. R. de Groot and P. Masur,
{\it Non-Equilibrium Thermodynamics},  (Dover, New York
1984).

\bibitem{liqCryst-1} P.G. de Gennes and J. Prost,
{\em The Physics of Liquid Crystals} (Clarendon Press,
Oxford 1993).
\bibitem{liqCryst-5} M. Liu,
{\it Hydrodynamic theory of biaxial nematics}, Phys. Rev.
{\bf A 24}, 2720 (1981).


\bibitem{he3-1} D. Vollhardt and P. W\"{o}lfle,
{\it The Superfluid Phases of Helium 3}, Taylor and
Francis, London (1990).
\bibitem{he3-3} M. Liu, {\it Hydrodynamics of $^3$He near the
A-Transition,} Phys. Rev. Lett. {\bf 35}, 1577 (1975).
\bibitem{he3-6} M. Liu, {\it Relative Broken Symmetry and the Dynamics of the
$A_1$-Phase,} Phys. Rev. Lett. {\bf 43}, 1740 (1979).


\bibitem{SC-1} M. Liu, {\it  Rotating Superconductors and the
Frame-independent London Equations,} Phys. Rev. Lett.
{\bf 81}, 3223, (1998).
\bibitem{SC-2} Jiang Y.M. and M. Liu, {\it
Rotating Superconductors and the London Moment:
Thermodynamics versus Microscopics,} Phys. Rev. {\bf B
6}, 184506, (2001).
\bibitem{SC-3} M.~Liu,  {\em Superconducting
Hydrodynamics and the Higgs Analogy,} J. Low Temp. Phys.
126, 911, (2002)

\bibitem{hymax-1} K. Henjes and M. Liu,
{\it Hydrodynamics of Polarizable Liquids,} Ann. Phys.
{\bf 223}, 243 (1993).
\bibitem{hymax-2} M. Liu, {\it Hydrodynamic Theory
of Electromagnetic Fields in Continuous Media,} Phys.
Rev. Lett. {\bf 70}, 3580 (1993).
\bibitem{hymax-4} Y.M.
Jiang and M. Liu, {\it Dynamics of Dispersive and
Nonlinear Media,} Phys. Rev. Lett. {\bf 77}, 1043,
(1996).

\bibitem{FF-2} R.E.
Rosensweig, {\em Ferrohydrodynamics}, (Dover, New York
1997).
\bibitem{FF-3} M. Liu, {\it Fluiddynamics of Colloidal Magnetic
and Electric Liquid,} Phys. Rev. Lett. {\bf 74}, 4535
(1995).
\bibitem{FF-8} O. M\"{u}ller, D. Hahn and M.
Liu, {\em Non-Newtonian behaviour in ferrofluids and
magnetization relaxation,} J. Phys.: Condens. Matter 18,
2623, (2006).
\bibitem{FF-9} S. Mahle, P. Ilg and M. Liu, {\em
Hydrodynamic theory of polydisperse chain-forming
ferrofluids,} Phys. Rev. {\bf E 77}, 016305 (2008).

\bibitem{polymer-1}
H. Temmen, H. Pleiner, M. Liu and H.R. Brand, {\it
Convective Nonlinearity in Non-Newtonian Fluids,} Phys.
Rev. Lett. {\bf 84}, 3228 (2000).
\bibitem{polymer-3} H. Pleiner, M. Liu and H.R. Brand,
{\it Nonlinear Fluid Dynamics Description of
non-Newtonian Fluids}, {Rheologica Acta} {\bf 43}, 502
(2004).
\bibitem{polymer-4}O. M\"{u}ller, {\em Die Hydrodynamische Theorie Polymerer Fluide}, PhD
Thesis University T\"{u}bingen (2006).

\bibitem{JL6}    Y.M. Jiang, M. Liu,
{\em Granular Solid Hydrodynamics}, Grannular
Matter,{\bf11-3}, 139 (2009) [DOI
10.1007/s10035-009-0137-3].


\bibitem{elaPla-1} R.M. Nedderman,
{\em Statics and Kinematics of Granular Materials}
(Cambridge University Press, Cambridge, 1992).
\bibitem{elaPla-2}A. Schofield, P. Wroth, {\em Critical State Soil
Mechanics} (McGraw-Hill, London, 1968).


\bibitem{Haff-1}  P. K. Haff,
{\em Grain flow as a fluid-mechanical phenomenon}, J.
\bibitem{Haff-2} J. T. Jenkins and S.
B. Savage, {\em A theory for the rapid flow of identical,
smooth, nearly elastic particles},  J. Fluid Mech.
\textbf{130}, 187(1983).

\bibitem{chute-2}GDR MiDi,
{\it On dense granular flows}, Eur. Phys. J. {\bf E 14},
341 (2004).
\bibitem{chute-3}P.Jop, Y. Forterre, O. Pouliquen, {\it A
constitutive law for dense granular flows},  Nature
{\bf441}, 727, 2006.

\bibitem{ge-1} D.O. Krimer, M. Pfitzner, K. Br\"{a}uer,
Y. Jiang, M. Liu, {\em Granular Elasticity: General
Considerations and the Stress Dip in Sand Piles,} Phys.
Rev. {\bf E74}, 061310 (2006).
\bibitem{ge-2}K. Br\"{a}uer, M. Pfitzner,
D.O. Krimer, M. Mayer, Y. Jiang, M. Liu, {\em Granular
Elasticity: Stress Distributions in Silos and under Point
Loads,} Phys. Rev. {\bf E74}, 061311 (2006);




\bibitem{J-L-1} Y.M. Jiang, M. Liu, {\em Granular Elasticity
without the Coulomb Condition,} Phys. Rev. Lett.
\textbf{91}, 144301 (2003).
\bibitem{J-L-2} Y.M. Jiang, M. Liu, {\it Energy Instability
Unjams Sand and Suspension,}  Phys. Rev. Lett.
\textbf{93}, 148001(2004).
\bibitem{J-L-3} Y.M. Jiang, M. Liu, {\em A Brief Review of
``Granular Elasticity",} Eur. Phys. J. {\bf E~22,} 255
(2007).
\bibitem{Lub-1} L. Bocquet, J. Errami, and T. C. Lubensky,
{\it Hydrodynamic Model for a Dynamical Jammed-to-Flowing
Transition in Gravity Driven Granular Media}, Phys. Rev.
Lett., \textbf{89}, 184301 (2002).
\bibitem{Lub-2} W. Losert, L.
Bocquet, T. C. Lubensky, and J. P. Gollub, {\it Particle
Dynamics in Sheared Granular Matter}, Phys. Rev. Lett.,
\textbf{85}, 1428 (2000);
\bibitem{Lub-3} L. Bocquet, W.
Losert, D. Schalk, T. C. Lubensky, and J. P. Gollub, {\it
Granular shear flow dynamics and forces: Experiment and
continuum theory},  Phys. Rev., E \textbf{65}, 011307
(2002);

\bibitem{JL3}  Y.M. Jiang, M. Liu,
{\em From Elasticity to Hypoplasticity: Dynamics of
Granular Solids,} Phys. Rev. Lett. {\bf 99}, 105501
(2007).
\bibitem{Houlsby}I. F. Collins and G. T. Houlsby,
{\em Application of thermomechanical principles to
the modelling of geotechnical materials}, Proc. R.
Soc. Lond. A  {\bf 453}, 1975, (1997).

\bibitem{Edw} S.F. Edwards, R.B.S. Oakeshott,
{\it Theory of powders}, Physica A {\bf 157}, 1080
(1989); S.F. Edwards, D.V. Grinev, {\it Statistical
Mechanics of Granular Materials: Stress Propagation and
Distribution of Contact Forces}, Granular Matter, {\bf
4}, 147 (2003).



\end{thebibliography}
\end{document}